\documentclass[amsmath,amssymb,prc,preprint,showpacs]{revtex4-2}

\usepackage{amssymb,amsmath} 
\usepackage{graphicx,epsfig} 

\newcommand{\Mg}{{}^{24}\mathrm{Mg}}
\newcommand {\mbf}[1]{{\mathbf{#1}}}
\newcommand {\vecg}[1]{\mbox{\boldmath{$#1$}} }
\newcommand{\cm}{\mathrm{c\!\:\!.m\!\:\!.}}

\begin{document}

\title{Nonlocal optical potential \\
  in the inelastic deuteron scattering off $^{24}$Mg}


\author{A. Deltuva}
\email{arnoldas.deltuva@tfai.vu.lt}
\affiliation
{Institute of Theoretical Physics and Astronomy, 
Vilnius University, Saul\.etekio al. 3, LT-10257 Vilnius, Lithuania
}

\author{D. Jur\v{c}iukonis}
\affiliation
{Institute of Theoretical Physics and Astronomy, 
Vilnius University, Saul\.etekio al. 3, LT-10257 Vilnius, Lithuania
}

\received{April 7, 2023}

\begin{abstract}%
  Nonlocal nucleon-nucleus optical   potential
  with rotational quadrupole deformation enabling the excitation of the ${}^{24}\mathrm{Mg}(2^+)$ state
   is developed; it fits well the 
  proton-${}^{24}\mathrm{Mg}${} elastic and inelastic differential cross section
  in the  beam energy range from 30 to 45 MeV per nucleon.
  The inelastic deuteron-${}^{24}\mathrm{Mg}${}
  scattering leading to the excited ${}^{24}\mathrm{Mg}(2^+)$ state is studied in the same energy regime
  by solving the three-body Faddeev-type equations for transition operators.
  Effects of the optical potential nonlocality are evaluated by comparison with local models.
  Significant effects  on the inelastic differential cross section
  are found at forward angles up to the first peak and at larger angles beyond the second peak.
  Nonlocal optical potential provides a simultaneous reasonable reproduction of the experimental data for 
the  elastic and inelastic proton-${}^{24}\mathrm{Mg}$
  and deuteron-${}^{24}\mathrm{Mg}$ scattering, not achieved using local potentials.
\end{abstract}

\maketitle

\section{Introduction}
In a recent study of deuteron stripping and pickup reactions \cite{deltuva:23a}
we united two important ingredients beyond the widely employed standard
nuclear dynamics, namely, the nonlocal extension of the nucleon-nucleus 
potentials and the excitation of the nuclear core. Furthermore,
we used rigorous three-body   Faddeev-type equations
for transition operators \cite{faddeev:60a,alt:67a}
and obtained accurate solutions in the momentum-space partial-wave framework.
The achieved description of the experimental data and the consistency
between the two-body  and three-body description
is considerably improved as compared to previous studies.

Encouraged by the above-mentioned success, in the present work
we aim to investigate the 
interplay of the optical potential nonlocality and 
the collective nuclear degrees of freedom, i.e., the nuclear core excitation (CeX), 
yet in another type of reactions, the inelastic deuteron scattering.
To the best of our knowledge, the importance of the nonlocality
in this reaction type is unexplored so far since it necessitates
the CeX which in previous studies was always assumed to have a local form.
We take the $\Mg${} nucleus as a working example, since (i) its lowest states
display rather well the rotational band structure describable by the
quadrupole deformation, (ii) the experimental differential cross section data are available
not only for elastic and inelastic scattering of deuterons \cite{mg:d},
but also for the elastic and inelastic nucleon-nucleus scattering \cite{pmgc},
which is necessary to constrain the potentials, and (iii) several quite
sophisticated
theoretical calculations using local potentials already exist. They are based on the extension
of the continuum discretized coupled channels (CDCC) method
\cite{chau:11a,gomez:17b} and the Faddeev-type theory \cite{deltuva:16a}.
As pointed out there, the two-body distorted-wave Born approximation (DWBA),
though able to fit the experimental data \cite{mg:d}, requires values for the
quadrupole deformation parameter $\beta_2$ that are inconsistent with
the  nucleon-nucleus data. The consistency was partially improved by using
three-body treatments such as CDCC or Faddeev, though definite conclusions were
precluded by the shortcomings of the used optical potentials.
The results in Refs.~\cite{gomez:17b,deltuva:16a} were obtained with
global optical potential parametrizations such as
Chapel Hill 89 (CH89) \cite{CH89} and Koning and Delaroche (KD) \cite{koning},
that provide reasonable but not perfect description of the nucleon-$\Mg${}
scattering data. One of our goals in the present work is the development 
of nonlocal optical potentials with an improved account of the two-body data. 
Furthermore, while some parameters of local optical potentials are energy dependent,
the nonlocal form exhibits a weaker energy dependence and therefore a good fit 
of the experimental data over a broader energy range is possible with energy-independent
parameters, thereby allowing to reduce the ambiguities and increase the predictive power
of the nonlocal optical potential \cite{deltuva:23a}.

In Section \ref{sec:eq} we recall the three-body Faddeev formalism, while
in Sec.~\ref{sec:pot} we develop nonlocal optical potentials and, for comparison, local ones.
In Section \ref{sec:mgd} we present analysis and discussion of the  deuteron-$\Mg${} scattering
observables, with conclusions summarized in Sec.~\ref{sec:con}.

\section{Three-particle equations with nuclear excitation \label{sec:eq}}

We consider the three-particle system of a proton ($p$), a neutron ($n$), and
a nuclear core with the mass number $A$, the latter being $\Mg${} in the considered case.
The version of three-particle Faddeev equations \cite{faddeev:60a}
for transition operators, 
directly related to scattering amplitudes,
was proposed by Alt, Grassberger, and Sandhas (AGS) \cite{alt:67a}. 
Extended to include the excitation of one of the particles, 
the nucleus $A$ in the present case, the AGS equations
for multi-component transition operators $U_{\gamma\beta}^{cb}$
read 
\begin{equation}  \label{eq:Uba}
U_{\gamma\beta}^{cb}  = \bar{\delta}_{\gamma\beta} \, \delta_{cb} {(G^{c}_{0})}^{-1}  +
\sum_{\alpha=p,n,A} \, \sum_{a=g,x}   \bar{\delta}_{\gamma\alpha} \, T_{\alpha}^{ca}  \,
G_{0}^{a} U_{\alpha \beta}^{ab}.
\end{equation}
Here the Latin superscripts label the internal states of the nucleus, either the
ground (g) state  $\Mg(0^+)$  or the excited (x) state $\Mg(2^+)$ with 1.369 MeV excitation energy
(we do not include higher excited states of  $\Mg$),
and the Greek subscripts label the spectator
particle in the odd-man-out notation, e.g., the spectator $p$ implies
the pair $A$+$n$, etc. 
Furthermore,  $\bar{\delta}_{\gamma\beta} = 1 - \delta_{\gamma\beta}$,
$G_{0}^{a}$ is the free resolvent in the respective Hilbert sector $a$,
and 
\begin{equation}  \label{eq:Tg}
T_{\alpha}^{ca} =  V_{\alpha}^{ca} +\sum_{b=g,x} 
V_{\alpha}^{cb} G_0^{b} T_{\alpha}^{ba}
\end{equation}
is the two-particle transition operator. For $A$+$n$ and $A$+$p$ pairs
it couples the two Hilbert sectors as the respective potential
$V_{\alpha}^{ca}$ does, since the nucleus $A$ can be excited/deexcited when
interacting with nucleons.

The amplitude for the deuteron inelastic scattering is determined
by the transition operator component $U_{AA}^{xg}$, which is, however, coupled
to the other five components  $U_{\gamma A}^{cg}$ via the AGS equations
(\ref{eq:Uba}). In Ref.~\cite{deltuva:16a}
all those equations for transition operator components
are given explicitly, together with the description of asymptotic channel
states, the Coulomb treatment via the screening and renormalization method,
and the relation to the differential cross section. The AGS equations are
solved in the momentum-space partial-wave representation, employing  three different
sets of basis states, as appropriate for the treatment of three interacting pairs of particles.
Further technical details of calculations are described 
in Ref.~\cite{deltuva:16a} and references therein.

\section{Nonlocal potential  \label{sec:pot}}

In our transition-operator integral-equation formalism the
nonlocal coordinate-space potentials $V_{\alpha}^{ca}$, transformed into the momentum space  representation,
do not require any special treatment as compared to local ones.
As in Ref.~\cite{deltuva:23a} we start with a single-particle nonlocal coordinate-space potential
 \begin{equation}  \label{eq:Vdj}
   \langle \mbf{r}' | V_N |\mbf{r} \rangle =
   \frac12 \big[ V(r') H(|\mbf{r}'-\mbf{r}|) + H(|\mbf{r}'-\mbf{r}|) V(r) \big]
\end{equation}
 where $\mbf{r}$ and $\mbf{r}'$ are initial and final distances between particles,
  $V(r)$ is local potential function of the respective distance, and
 \begin{equation}  \label{eq:Hx}
 H({x}) = \pi^{-3/2} \rho^{-3} e^{-(x/\rho)^2}
\end{equation}
 is the nonlocality function with the nonlocality range $\rho$.
 As argued in Ref.~\cite{deltuva:23a}, this phenomenological form is closely related to the 
  Perey and Buck potential \cite{pereybuck},
  and in the limit $\rho \to 0$  one obviously recovers the local potential $V(r)$.
  We parametrize  $V(r)$ in the same way as done for standard optical potentials
  \cite{CH89,koning}, i.e.,
\begin{equation}  \label{eq:Vr}
   V(r) =  -V_V \,f_V(r) - iW_V \,f_W(r) - i 4 W_S\,f_S(r)[1-f_S(r)] 
   + V_s \frac{2}{r} \frac{df_s(r)}{dr} \, \vecg{\sigma}\cdot \mbf{L}.
 \end{equation}
The four terms with strength parameters $V_V$, $W_V$, $W_S$, and  $V_s$,
correspond to the real volume, imaginary volume, imaginary surface,
and real spin-orbit contributions, respectively. Their radial dependence
is modeled by the standard  Woods-Saxon functions
 \begin{equation}  \label{eq:fws}
   f_j(r) =  \frac{1}{1+ e^{(r-R_j)/a_j}}
\end{equation}
 with parametric dependence on the radius $R_j$ and diffuseness $a_j$.

 The potential given by Eqs.~(\ref{eq:Vdj}) - (\ref{eq:fws})
 does not act on internal degrees of freedom of the nucleus $A$ and therefore
 does not induce its excitation/deexcitation. This can be achieved by its deformation
 \cite{bohr-motelson,tamura:cex,thompson:88}.
 The rotational model, quite consistent with the low-energy spectrum 
 of $\Mg${}, assumes a permanent quadrupole deformation of $\Mg${}. This effectively
 results in the  Woods-Saxon radius $R_j = R_{j0}[1+\beta_2 Y_{20}(\hat{\xi})]$,
 where  $\beta_2$ is the quadrupole deformation parameter and $\hat{\xi}$
 describes the internal nuclear degrees of  freedom in the body-fixed frame
 \cite{tamura:cex,thompson:88,deltuva:16a}. For calculations in the partial-wave representation
 the deformed  potential has to be expanded  into multipoles.
In addition to the central contribution $\lambda=0$ one has to include also the 
$\lambda=2$ multipole to induce the transitions between $0^+$ and $2^+$ states of $\Mg${}
\cite{gomez:17b,tamura:cex,thompson:88,deltuva:16a}.
In the coordinate-space partial-wave basis $ |rLSJ \rangle$, where
$L$, $S$ and $J$ denote the two-particle relative orbital momentum,
the total spin $S$,  and the conserved total angular momentum $J$, respectively,
the two functions $H(|\mbf{r}'-\mbf{r}|)$ and $ V(r)$
are transformed separately, and then combined into a nonlocal potential
\begin{equation}  \label{eq:Vdjl}
    \langle r'L'S'J| V_{N}^{ca} |rLSJ \rangle =  \frac12 \Big[  V_{L'S',LS,J}^{ca}(r') H_{L}(r',r)
    + H_{L'}(r',r) V_{L'S',LS,J}^{ca}(r)  \Big].
\end{equation}
Here 
\begin{equation}  \label{eq:HL}
  H_{L}(r',r) = 2\pi \int_{-1}^1 dx P_L(x) \, H(\sqrt{{r'}^2+r^2-2r'rx})
\end{equation}
is the partial-wave projection of $H(|\mbf{r}'-\mbf{r}|)$, with  $P_L(x)$ being the Legendre polynomial,
and 
$V_{L'S',LS,J}^{ca}(r)$ is the  standard local potential with the CeX
\cite{tamura:cex,thompson:88}, whose $c \neq a$ components arise from the $\lambda=2$ multipole
and couple different internal states of the nucleus $A$. A local potential of this type has been
used also in  previous calculations of deuteron-$\Mg${} scattering \cite{gomez:17b,deltuva:16a}.

Finally, the transformation of the potential (\ref{eq:Vdjl}) to the momentum space,
as required for our calculations, is straightforward, i.e.,
\begin{equation}  \label{eq:Vp}
  \langle p'L'S'J| V_{N}^{ca} |pLSJ \rangle = (-1)^{\frac{L'-L}{2}} \, \frac{2}{\pi}
  \int_0^\infty  dr' \, dr \, {r'}^2 r^2 \, j_{L'}(p'r') 
    \langle r'L'S'J| V_{N}^{ca} |rLSJ \rangle j_{L}(pr),
\end{equation}
where $p$ is the relative nucleon-nucleus momentum, and $j_L(x)$ is the spherical Bessel function
of the order $L$.
For the proton-nucleus interaction the Coulomb contribution, including the deformation
\cite{tamura:cex,deltuva:16a}, is added.

The values of the optical potential parameters have to be obtained by fitting the experimental data.
We are interested in deuteron-$\Mg${} scattering at  30 to 45 MeV/nucleon beam energies  \cite{mg:d},
and the elastic and inelastic $p+\Mg${} data  in this energy regime  is
available  \cite{pmgc}. The data for the $n+\Mg${} reaction are quite scarce and only  available at lower
energies. On the other hand, $\Mg$ is an isospin symmetric $Z=N$ nucleus, consequently, after
separation of the Coulomb contributions, the nuclear parts of the
proton and neutron optical potentials  should be similar, as it is the case for global parametrizations 
\cite{CH89,koning,giannini}. We therefore take  the optical potential parameters for the
neutron-nucleus to be those for the proton-nucleus.

One could consider the strengths, radii and diffuseness for the four terms in Eq.~(\ref{eq:Vr})
plus $\beta_2$ and $\rho$ as fit parameters. However, for a fair comparison with previous results
we demand that the number of free parameters in our potential does not exceed the one in 
standard optical potentials, and introduce additional constrains:
(i) the nonlocality range is fixed to $\rho=1$ fm,  a typical value for nonlocal
potentials \cite{deltuva:23a,giannini,giannini2};
(ii) geometric parameters for both volume and surface imaginary terms are chosen to be the same,
i.e., $R_W=R_S$ and $a_W=a_S$, as it is often the case for standard parametrizations  \cite{CH89};
(iii) since polarization observables are not studied, and cross sections are insensitive to the
spin-orbit interaction, it is not subjected to the fit and not deformed. Instead, we
assume $R_s=R_{V0}$,  $a_s=a_V$ and $V_s = 7.5\, \mathrm{MeV\,fm^2}$, consistently
with the polarization data for ${}^{16}\mathrm{O}$ \cite{deltuva:23a}.
With these constrains our fitting parameters are $\beta_2$,
$V_V$, $W_W$, $W_S$, $a_V$, $a_W$, $r_V$ and $r_W$, where the reduced radii $r_j$ are
related to Woods-Saxon radii in the standard  way $R_{j0} = r_j A^{1/3}$.

\begin{table}[!h]
  \caption{Five example sets for nonlocal  $p+\Mg${} optical potential parameters
    with $\rho=1$ fm.
The strengths  $V_V$, $W_W$ and $W_S$ are in MeV, while $r_j$, $a_j$ and $\delta_2$ are in fm.
}
\label{tab:v}
\centering
\begin{ruledtabular}
\begin{tabular}{*{9}{r}}
  $V_V$ & $W_W$ & $W_S$ & $r_V$ &  $r_W$ & $a_V$  & $a_W$ & $\beta_2$ &  $\delta_2$ \\ \hline
 106.64 &  7.76 & 6.84 & 1.01 & 1.10 & 0.70 & 0.67 & 0.54 & 1.57 \\ 
 100.75 &  7.25 & 9.80 & 1.07 & 1.09 & 0.66 & 0.56 & 0.50 & 1.56 \\ 
 101.28 &  8.01 & 9.73 & 1.05 & 1.03 & 0.67 & 0.61 & 0.52 & 1.60 \\ 
 105.34 &  7.46 & 7.37 & 1.02 & 1.04 & 0.68 & 0.75 & 0.53 & 1.58 \\ 
 108.79 &  7.39 & 8.66 & 0.99 & 1.03 & 0.69 & 0.70 & 0.57 & 1.64 \\ 
\end{tabular}
\end{ruledtabular}
\end{table}

\begin{figure}[!h]
  \centering\includegraphics[scale=0.8]{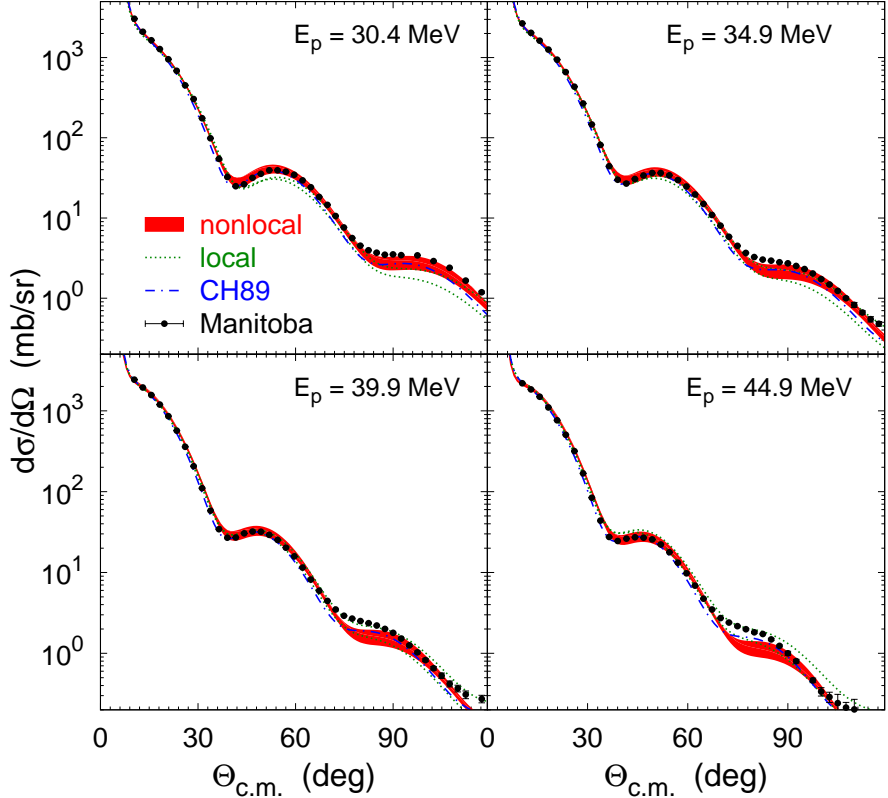}
\caption{Differential cross section for elastic $p+\Mg${} scattering at
  beam energies $E_p = 30.4$, 34.9, 39.9 and 44.9 MeV.
  Results obtained with different parameter sets of the nonlocal optical potential
  are combined into bands, while  curves represent predictions
  based on  local potentials, i.e., those developed in this work
  (dotted) and CH89 (dashed-dotted). The experimental data 
  from Ref.~\cite{pmgc} were  measured at University of Manitoba.}
\label{fig:p}
\end{figure}

\begin{figure}[!h]
  \centering\includegraphics[scale=0.8]{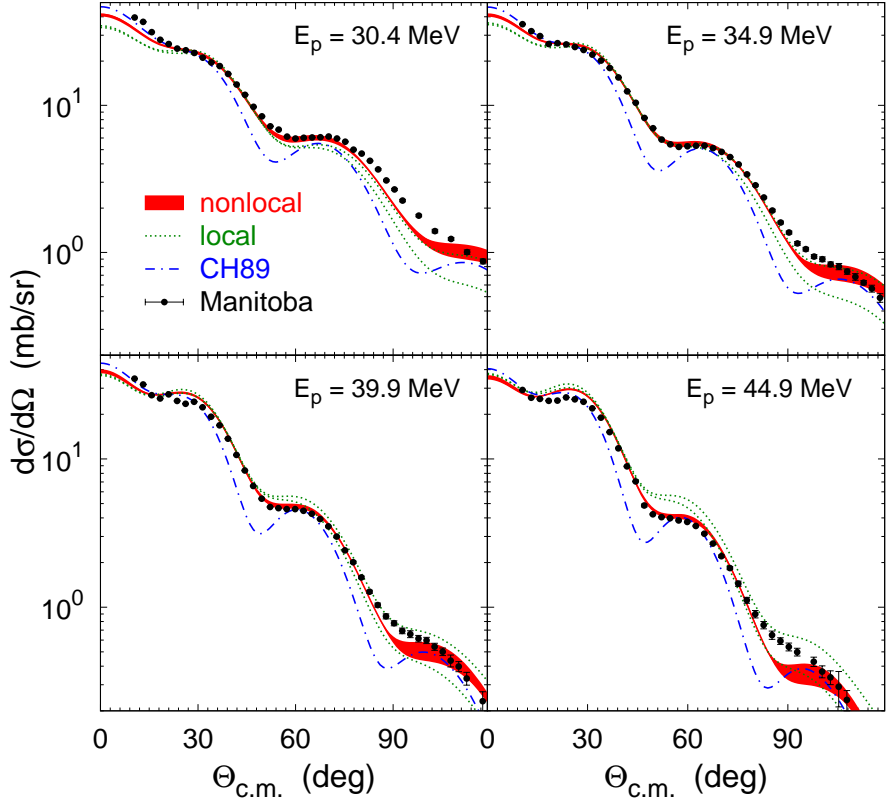}
  \caption{Differential cross section for inelastic $p+\Mg${} scattering leading to the $\Mg(2^+)$ state.
    Results  at   beam energies $E_p = 30.4$, 34.9, 39.9 and 44.9 MeV are shown.
    Bands, curves and experimental data are as in Fig.~\ref{fig:p}.}
\label{fig:px}
\end{figure}

We fit simultaneously differential cross section for
elastic and inelastic $p+\Mg${} scattering at
beam energies $E_p = 30.4$, 34.9, 39.9 and 44.9 MeV \cite{pmgc}. In order to achieve a better fit
at smaller angles, relevant for the deuteron scattering, we exclude  the data at center-of-mass angles
$\Theta_\cm$ beyond 70 deg. We  estimate the uncertainties by  developing a number of parameter sets
that fit the data with a comparable quality; several typical examples are collected in Table
\ref{tab:v}. The quadrupole deformation parameter $\beta_2$ takes the values from 0.50 to 0.57, i.e.,
with spread of nearly 15\%,
but the deformation length $\delta_2 = \beta_2 R_{V0}$ varies within 5\% only, ranging from
1.56 to 1.64 fm.
The corresponding predictions for $p+\Mg${} elastic
and inelastic scattering to the $2^+$ state are displayed as  bands in
Figs.~\ref{fig:p} and \ref{fig:px}, and  agree reasonably well with the experimental data \cite{pmgc}.

In order to evaluate importance of the optical potential nonlocality,
we attempted to fit the same data with $\rho=0$, i.e., using an energy-independent
local potential. The local model has the same constrains as nonlocal one, except
that $V_s = 6.0\, \mathrm{MeV\,fm^2}$, a typical value for local potentials.
As expected, such a potential is less successful in a broader energy range,
the deviations from the data are most evident for the lowest and highest considered energy.
For each observable we present in Figs.~\ref{fig:p} and \ref{fig:px}
two dotted curves, their difference may provide some
estimation of uncertainties. The deduced values for $\beta_2 = 0.52$ and 0.55  
are consistent with nonlocal cases, while for $\delta_2 = 1.73$ and 1.77 fm they are slightly larger. 
In addition, as dashed-dotted curves we include the predictions based on
the global optical potential parametrization CH89 \cite{CH89}, deformed with $\beta_2=0.5$
and $\delta_2 = 1.69$ fm and already used in previous studies \cite{gomez:17b,deltuva:16a}.
This potential 
has energy-dependent parameters. It reproduces reasonably well the elastic differential cross section
but fails for
inelastic data at $\Theta_\cm > 40$ deg, having a different shape of the angular distribution.
Thus, the optical potentials developed in the present work yield a significant improvement in the
description of inelastic $p+\Mg${} scattering.

\section{Results for three-body reaction  \label{sec:mgd}}

We proceed to the analysis of $\Mg(d,d')$ differential cross sections at deuteron
beam energies $E_d = 60$, 70, 80 and 90 MeV, as measured at J\"ulich Research Center \cite{mg:d}.
Solution of the  three-body AGS equations with the quadrupole excitation of the $\Mg${} nucleus
requires three pair potentials as input. The proton-nucleus and neutron-nucleus optical potentials
are taken from the previous section; as there, the proton-nucleus interaction is appended with
central and deformed Coulomb terms. The neutron-proton interaction is modeled with
a realistic CD Bonn potential \cite{machleidt:01a}.

\begin{figure}[!h]
  \centering\includegraphics[scale=0.8]{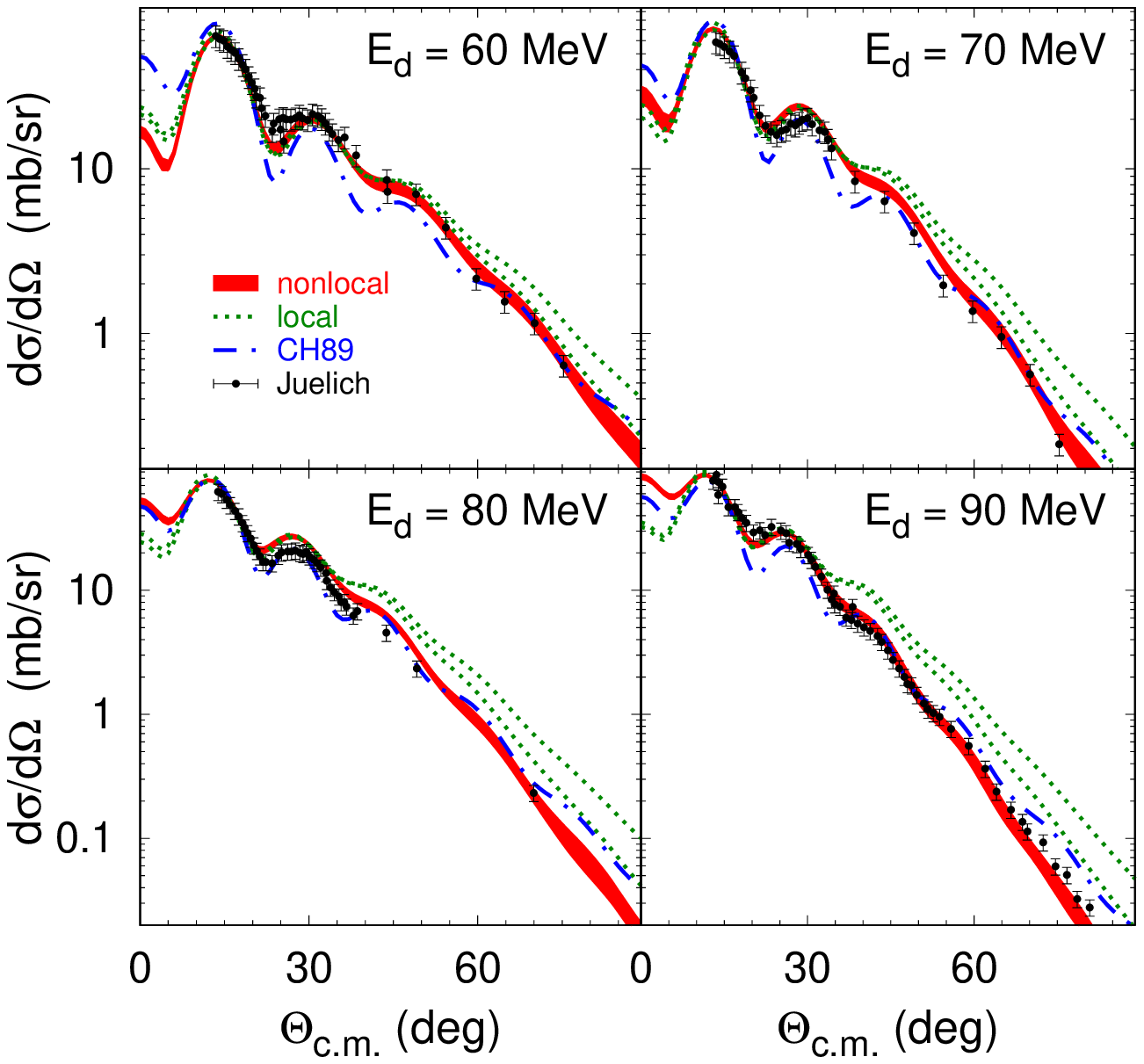}
\caption{Differential cross section for inelastic $d+\Mg${} scattering leading to the $\Mg(2^+)$ state.
    Results  at deuteron  beam energies $E_d = 60$, 70, 80 and 90 MeV are shown.
    Bands and  curves  are as in Fig.~\ref{fig:p}. The experimental data 
  from Ref.~\cite{mg:d} were  measured at J\"ulich Research Center.}
\label{fig:dx}
\end{figure}

In Fig.~\ref{fig:dx} we present the calculated differential cross sections for the $\Mg(d,d')$ reaction
with the final $\Mg${} nucleus being in its first excited state $2^+$.
The predictions obtained with nonlocal potentials are displayed as bands,
those with local potentials as dotted curves, and those based on the CH89 are
shown as dashed-dotted curves. The comparison of different calculations and experimental data
lead to the following important observations:

(i) The CH89-based differential cross sections exhibit minima near $\Theta_\cm = 20 $ and 40 deg that
are considerably deeper than those seen in data and other calculations. This is likely  a consequence
of a similar discrepancy seen in the  $\Mg(p,p')$ inelastic scattering in Fig.~\ref{fig:px}
near  $\Theta_\cm = 50 $ and 90 deg. Thus, the failure of the CH89 model around the minima can be explained
by the shortcomings present already at the two-body level.

(ii) The first peak near $\Theta_\cm = 15$ deg  is best reproduced by the band of nonlocal optical potentials,
with slight overprediction at 70 and 80 MeV. The local models, especially CH89 at 60 and 70 MeV, show larger
overprediction. This difference has no evident explanation at the two-body level.

(iii) Both local and nonlocal models, fitted in this work to proton-nucleus data, mutually agree quite well around
the second peak, that moves from 30 to 25 deg with increasing beam energy.
The agreement with data is quite good at 60 and 90 MeV, but overprediction up to 20\% is observed  at 70 and 80 MeV.
In contrast, the CH89 provides a better agreement at 70 and 80 MeV, but underpredicts the data at 60 and 90 MeV.
The energy evolution is smooth for the predictions but not for the data points,
raising some concerns regarding their  accuracy.

(iv) Predictions using nonlocal optical potentials describe the experimental data well also at large scattering
angles while local models clearly overpredict the data. This again finds no explanation by looking back into
the nucleon-nucleus scattering, as there the local model predictions may be even below those of nonlocal models
and data, as happens in  Fig.~\ref{fig:px} at $E_p=30.4$ MeV.

(v) All three types of employed models predict different energy evolution of the differential cross section
at forward angles  $\Theta_\cm < 10$ deg below the first peak, with no evident explanation at the two-body level.
Unfortunately, no experimental data is available in this angular regime.

We also would like to discuss several uncertainties related to the above study and argue that the main conclusions
remain unaffected.

(a) Changing the nonlocality parameter $\rho$ by $\pm 10$\%
and refitting other parameters  does not change visibly the description of two- and three-body data
shown in Figs.~\ref{fig:p} - \ref{fig:dx} for $\rho = 1$ fm. Changes  in the spin-orbit force only become
visible at large angles, beyond the scale of  Figs.~\ref{fig:p} - \ref{fig:dx}.

(b) At $E_p=30$ MeV we developed single-energy local potential fitting proton-$\Mg${} data as good as
nonlocal models do. Nevertheless,
the three-body results follow closely the trend of local models in  Fig.~\ref{fig:px}

(c) Since global optical potentials such as CH89 have both proton-nucleus and neutron-nucleus parametrizations,
we verified the effect of using  proton-nucleus parameters for the  neutron-nucleus pair.
The effect turns out to be visible only at very small angles, reaching 5\% at $\Theta_\cm = 0$ deg but decreasing
to roughly 1\% at the first peak and beyond it.

(d) We included only the first excited state $2^+$ of $\Mg${}, while the CDCC-type calculations \cite{gomez:17b}
investigated the effect of the second excited state $4^+$ and found up to 7\% reduction of the $\Mg(d,d')$
differential cross section, most visible at the first peak.
However, one has to keep in mind that the inclusion of the  $4^+$ state changes also the $\Mg(p,p')$ predictions.
In principle, one should refit the potential parameters, otherwise also the $\Mg(p,p')$ cross section is slightly
decreased and the observed effect in the  $\Mg(d,d')$ reaction is partially caused by changes in $\Mg(p,p')$.
As the refitting was not performed in Ref.~\cite{gomez:17b},
the real effect of the  $4^+$ state is expected to be less significant.

\begin{figure}[!h]
  \centering\includegraphics[scale=0.8]{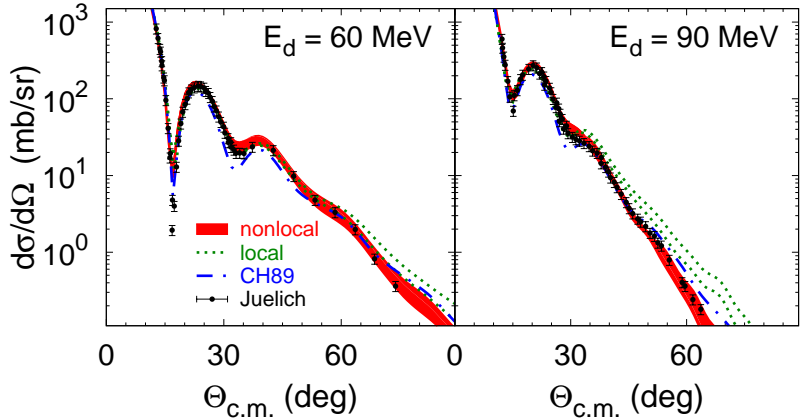}
\caption{Differential cross section for elastic $d+\Mg${} scattering at 60 and 90 MeV deuteron beam energy.
    Bands,  curves and experimental data  are as in Fig.~\ref{fig:dx}.}
\label{fig:d}
\end{figure}

The nonlocality effect in the deuteron-nucleus elastic scattering has been investigated previously
\cite{deltuva:09b},  as it does not demand including the nuclear excitation. 
Therefore we show in Fig.~\ref{fig:d} the $\Mg(d,d)$ differential cross section at
lowest and highest considered energy only. Despite different nucleus and more elaborated optical potentials
of the present work, the nonlocality effect appears to be qualitatively similar to the one observed
in Ref.~\cite{deltuva:09b} for the elastic deuteron scattering off ${}^{16}\mathrm{O}$ and ${}^{40}\mathrm{Ca}$ nuclei.
As compared to local models, the
nonlocal ones predict lower differential cross section at intermediate and large angles,
and this change is clearly favored by the experimental data, as can be seen in Fig.~\ref{fig:d} as well.

\section{Conclusions \label{sec:con}}

We studied the effect of the optical potential nonlocality in the inelastic deuteron-nucleus scattering.
$\Mg${} nucleus with the excited $2^+$ state was chosen as a working example. We developed a nonlocal nucleon-nucleus optical
potential with rotational quadrupole deformation, coupling ground and excited states, 
and fitted to proton-$\Mg${} elastic and inelastic differential cross section. A good reproduction of data
in the 30 to 45 MeV range was achieved with energy-independent parameters; several parameter sets
were determined to estimate the uncertainties. The local models, especially
the CH89 potential used in earlier calculations, are less successful in reproducing  experimental proton-$\Mg${} data.

We described the elastic and inelastic deuteron-nucleus scattering  using rigorous three-body Faddeev-type equations
for transition operators, and solved them in the momentum-space partial-wave representation.
Differential cross sections were calculated at  deuteron beam energies 60 to 90 MeV.
Using nonlocal models we obtained a good description of the inelastic experimental data, the overprediction of the second
peak at 70 and 80 MeV  may point to the inconsistency of the data sets at different energies.
The results based on the global CH89 potential fail in the minima regions, which can be
explained by the shortcomings of the potential in the nucleon-nucleus system.
The most visible effects of the optical potential nonlocality occur at forward angles up to the first peak
and at larger angles beyond the second peak. While in the former case the experimental data are not available,
in the latter case
nonlocal models are clearly favored  over the local ones. These differences have no evident explanation
in the nucleon-nucleus system. We also argued that simplification assumptions such as
using the same parameters for proton and neutron
optical potentials and neglecting  higher excited states are not expected to change the conclusions.

Finally, the nonlocality effect in the 
elastic deuteron-nucleus scattering is consistent with previous studies of other nuclei.
It reduces the   differential cross section at larger angles improving the agreement with the experimental data.

In summary, using nonlocal optical potentials we obtained a simultaneous satisfactory reproduction of
the experimental data for 
elastic and inelastic proton-$\Mg${} and deuteron-$\Mg${} scattering, not achieved in previous studies.
 A new measurement at forward angles could provide even more stringent test.

\vspace{1mm}
\begin{acknowledgments}
This work was supported by Lietuvos Mokslo Taryba
(Research Council of Lithuania) under Contract No.~S-MIP-22-72.
\end{acknowledgments}


 \end{document}